\documentclass[runningheads]{llncs}
\usepackage[T1]{fontenc}
\usepackage[export]{adjustbox}

\usepackage{graphicx}
\usepackage{caption}
\usepackage{amsmath}
\usepackage{tikz}
\usetikzlibrary{calc}
\usepackage{enumitem}
\newcommand\T[1]{\noindent\vspace{0pt}\textbf{#1} }
\usepackage{url}

\usepackage{breakurl}
\usepackage{supertabular}
\usepackage{booktabs}
\usepackage{pifont}
\usepackage{flushend}
\usepackage{enumitem}

\usepackage{subcaption}
\usepackage[subtle,wordspacing=normal, tracking=normal]{savetrees}

\makeatletter
\renewcommand\section{\@startsection{section}{1}{\z@}%
                       {-8\p@ \@plus -4\p@ \@minus -4\p@}%
                       {6\p@ \@plus 4\p@ \@minus 4\p@}%
                       {\normalfont\large\bfseries\boldmath
                        \rightskip=\z@ \@plus 8em\pretolerance=10000 }}
\renewcommand\subsection{\@startsection{subsection}{2}{\z@}%
                       {-8\p@ \@plus -4\p@ \@minus -4\p@}%
                       {6\p@ \@plus 4\p@ \@minus 4\p@}%
                       {\normalfont\normalsize\bfseries\boldmath
                        \rightskip=\z@ \@plus 8em\pretolerance=10000 }}
\renewcommand\subsubsection{\@startsection{subsubsection}{3}{\z@}%
                       {-4\p@ \@plus -4\p@ \@minus -4\p@}%
                       {-1.5em \@plus -0.22em \@minus -0.1em}%
                       {\normalfont\normalsize\bfseries\boldmath}}
\makeatother

\begin{document}

\title{Ethereum Proof-of-Stake Consensus Layer:\\ Participation and Decentralization}
\titlerunning{Ethereum Proof-of-Stake Consensus Layer}

\author{Dominic Grandjean, Lioba Heimbach\thanks{corresponding author} and Roger Wattenhofer\\\email{\{grdomini,hlioba,wattenhofer\}@ethz.ch} }
\institute{ ETH Zurich}

\authorrunning{D. Grandjean, L. Heimbach and R. Wattenhofer}
%
%
\institute{ETH Zurich}
\maketitle              
\begin{abstract}
In September 2022, Ethereum transitioned from  \textit{Proof-of-Work (PoW)} to \textit{Proof-of-Stake (PoS)} during ``the merge'' --- making it the largest PoS cryptocurrency in terms of market capitalization. With this work, we present a comprehensive measurement study of the current state of the Ethereum PoS consensus layer on the \textit{beacon chain}. We perform a longitudinal study of the history of the beacon chain. Our work finds that all dips in network participation are caused by network upgrades, issues with major consensus clients, or issues with service operators controlling a large number of validators. Further, our longitudinal staking power decentralization analysis reveals that Ethereum PoS fairs similarly to its PoW counterpart in terms of decentralization and exhibits the immense impact of (liquid) staking services on staking power decentralization. Finally, we highlight the heightened security concerns in Ethereum PoS caused by high degrees of centralization.
\end{abstract}

\section{Introduction}\label{sec:intro}

The global market capitalization of cryptocurrencies currently exceeds a staggering US\$~1T~\cite{CoinMarketCap2023}. This value is secured by nodes in various open \textit{peer-to-peer (P2P)} networks. These nodes follow the consensus protocol to record and verify transactions. The \textit{decentralization}, i.e., fragmentation of control, of the node network is fundamental. Decentralization ensures that a small number of entities cannot manipulate the blockchain's records. Moreover, enhancing the decentralization of the consensus layer enhances censorship resistance, as no single party can exert significant control over the inclusion of transactions on the ledger.

To safeguard the distributed network from Sybil attacks, where a party tries to gain an advantage by creating numerous nodes, most blockchains employ either \textit{Proof-of-Work (PoW)} or \textit{Proof-of-Stake (PoS)} mechanisms. In PoW \textit{miners} solve computational puzzles, while in PoS \textit{validators} must stake (lock) the cryptocurrency's native token. In September 2022, Ethereum switched from PoW to PoS during ``the merge''. Ethereum is the second biggest cryptocurrency by market capitalization~\cite{CoinMarketCap2023} and the largest PoS cryptocurrency. With its move from PoW to PoS, Ethereum aimed to reduce energy usage but also increase decentralization by lowering the entry barriers for network participants~\cite{PoSvsPoWEthereum2023}.

In Ethereum PoS, network participants wishing to partake in the consensus --- be a validator --- must deposit 32~ETH. Validators have three tasks: first, continuously attest to the validity of the blocks created by other validators, second, participate in sync committees, and third, occasionally propose a block. In contrast to PoW, PoS requires continuous active participation of more than two-thirds of the validators for the blockchain to make progress. Therefore, for a PoS blockchain, it is not only essential that the consensus layer is decentralized but also crucial that its validators participate actively even when it is not their turn to propose a block.

Additionally, the switch from PoW to PoS introduced additional consensus layer security concerns. \textit{Maximal extractable value (MEV)}, which refers to the maximum value that can be extracted through including, excluding, and re-ordering the transactions in a block, was prevalent in Ethereum PoW with more than US\$~675M extracted value before the merge~\cite{FlashbotsPreMergeData2023}. Thus, MEV poses consensus layer security concerns, as it incentivizes rational miners to fork the blockchain~\cite{QinQuantifying2022}. Not only does MEV remain a concern in Ethereum PoS, but there are also new types of MEV opportunities, e.g., \textit{multi-block MEV}, that arise as a result of knowing the block proposer minutes in advance~\cite{MackingaTWAP2022}.

\T{Contributions.} In this work, we present the first comprehensive measurement study on the participation level and decentralization of the Ethereum PoS consensus layer. We summarise our contributions as follows: 
\begin{itemize}[topsep=0pt,itemsep=0pt]
    \item We study the participation level of validators in the Ethereum PoS and find that the participation levels are very high --- exceeding 98\%. Dips in participation levels generally coincide with network upgrades or bugs in one or more consensus clients. Additionally, we only find very few slashable offenses, i.e., instances of equivocation by a validator.  
    \item To investigate the decentralization of the validator landscape, we cluster Ethereum validators into entities to find that the level of decentralization of the Ethereum consensus layer has not significantly increased since the merge.
    \item We highlight the challenges of increasing consensus layer decentralization, i.e., incentivizing users to bypass large staking services but to run their own validators. Large entities do not only receive higher consensus layer rewards, as their participation levels are higher, but they also have the unique opportunity to extract multi-block MEV and thereby are expected to also have higher execution layer rewards.
\end{itemize}

\section{Ethereum Proof-of-Stake}\label{sec:background}
During the merge on 15 September 2022~\cite{EthereumMerge2023}, Ethereum transitioned from PoW to PoS as Sybil resistance. Ethereum now runs two layers: the execution and consensus layer. The execution layer, resembling the previous PoW protocol, retains the responsibility of validating and executing transactions. On the other hand, the consensus layer, constructed atop the beacon chain, focuses on achieving consensus among validators. Importantly, in the PoS paradigm, participants known as validators have replaced traditional miners. Validators are responsible for proposing and validating blocks in the Ethereum network. To become a validator, one must \textit{stake} (i.e., lock) 32~ETH into the designated deposit contract.

\subsection{Block Generation}

In contrast to PoW, where the timing of blocks is dictated by mining difficulty, PoS operates with a fixed tempo for block generation. To be precise, time is split into epochs. An epoch represents a fixed period in the Ethereum network, consisting of 32 slots. Each slot, in turn, is a time interval during which a single block can be proposed and validated. The duration of a slot is fixed at 12 seconds~\cite{KashyapPoSEthereum2023}, i.e., block production is synchronous.

As previously mentioned, there is a chance for a single block to be added to the blockchain in every slot. As used to be the case with PoW Ethereum, a block contains a collection of transactions~\cite{SmithBlocksEthereum2023}. In each slot, a validator is selected pseudo-randomly as the block proposer. If a slot's proposer fails to propose a block within the allotted time, the slot remains empty. We note here that the probability of a validator being chosen as a proposer is inversely proportional to the number of active validators in the network at the time of selection~\cite{ConsecutiveBlocks203}. In addition to being tasked with block proposals, validators are also assigned to committees to validate newly proposed blocks. We note here that validators chosen to propose or validate blocks are determined at least one epoch and at most two epochs in advance~\cite{VitalikEthereumSpec2020}. Finally, validators participate in sync committees to allow light clients to determine the head of the beacon chain.

\subsection{Validator Duties}
In the following, we detail the tasks performed by Ethereum PoS validators. Besides having to deposit 32~ETH, a validator must also operate three distinct software components: an execution client, a consensus client, and a validator client. Following the deposit, users enter an activation queue, which serves to control the influx of new validators joining the network. Once a validator joins the network, they are assigned three primary tasks:

\T{Block proposal.} Validators are sporadically selected as a block's proposer, which involves proposing new blocks and making them available for attestation. A pseudo-random selection process for block proposers ensures a fair distribution of block proposal opportunities among all validators. At the present state of the network, an individual validator typically gets an opportunity to propose a block approximately every 2.5 months.

\T{Block attestation.} Attestation involves validators confirming the validity and accuracy of the data contained within a block. Validators are expected to attest to their view of the head of the beacon chain, the most recent fully validated block, once per epoch. During each epoch, every validator submits an attestation to indicate their opinion on the head of the chain. Note that occasionally, validators are assigned the task of aggregating attestations from other validators in the same committee.

\T{Sync committee participation.} Sync committees have a duration of 27 hours and validators are pseudo-randomly selected to participate in a sync committee~\cite{EthereumSpec2022}. A sync committee creates signatures to attest to the chain's head that can be used by light clients to determine the head of the beacon chain.

\subsection{Validator Rewards and Penalties}\label{sec:rewardandpenality}
Validators receive rewards for the previously outlined tasks. Their rewards can be divided into consensus and execution layer rewards. Note that consensus layer rewards were received by validators since the start of the beacon chain in December 2020, while execution layer rewards only became available to validators after the merge, i.e., when PoS replaced PoW on Ethereum.

\T{Consensus layer rewards.} Validators receive rewards for block proposal, attestation (i.e., attesting to the source epoch, target epoch, and chain head), and participation in sync committees~\cite{EthereumSpec2022}. Additionally, validators receive whistle-blowing rewards for providing evidence of dishonest validators. Consensus layer rewards decrease on an individual validator basis as more validators join the network. Currently, a validator receives approximately 0.04~ETH for a successful proposal and 0.00001~ETH for a successful attestation~\cite{EdgingtonTechnicalOverview2023}.

\T{Execution layer rewards.} A block proposer also receives rewards from priority fees and direct user payments. These rewards are consensus layer rewards, which were introduced after the merge. On average, the execution layer reward of a proposal is around 0.1~ETH per block~\cite{ThalmanEthereumRewards2023}.

\T{Penalties and slashing.} To incentivize network participation and honest behavior, validators receive penalties for missing target and source votes. These penalties are equal in amount to the rewards received for successful attestation. Additionally, validators can also be slashed for serious offenses (e.g., proposing and signing two different blocks for the same slot). Slashing removes at least 1/32 of a validator's staked Ether.

\subsection{Staking Services}

Staking services give users easier access to Ethereum staking. Generally, staking services are either \textit{custodial}, where the service holds the user's keys, or \textit{non-custodial}, where the user retains control of the keys. Liquid staking further offers tokenized representations of staked assets.

\T{Custodial staking services.}
Custodial staking services, e.g., Binance, Bitcoin Suisse, Coinbase, and Kraken, hold the user's private keys and manage the technical aspects of staking. Users gain convenience, but besides paying a fee, they must also place trust in the service provider to safeguard their assets.

Further, custodial staking services such as Lido and Rocket Pool are staking pools governed by on-chain communities. In Lido, 30 permissioned companies provide staking services to the protocol, while Rocket Pool employs over 2,500 permissionless node operators for staking user funds.

\T{Non-custodial staking services.}
Non-custodial staking services, such as Stakefish and Staked, provide the infrastructure for staking but allow users to keep control of their assets. The service runs the validator nodes, but users interact directly with smart contracts to stake their assets.

\section{Data Collection}\label{sec:data}
We collect Ethereum execution and consensus layer data by running a Lighthouse consensus client and an Erigon execution client. Our consensus layer data set covers the period from the genesis of the beacon chain on 1 December 2020 (i.e., slot 0) through 15 May 2023 (i.e., slot 6,447,598). Notice that the beacon chain launched well in advance of the merge. In the time before the merge, the beacon chain was reaching consensus on its state without processing mainnet transactions~\cite{EthereumMerge2023}. Additionally, our execution layer data covers the period from the genesis of Ethereum on 30 July 2015 through 15 Mai 2023 (i.e., block 17,268,587). We further enhance our data set with validator and address labels from the Rated Network API~\cite{Rated2023}, \texttt{beaconcha.in}~\cite{Beaconchain2023}, and Etherscan~\cite{Etherscan2023} to allow for validator clustering. Appendix~\ref{app:datacollection} provides a detailed overview of the data collection.

\section{Beacon Chain Participation}\label{sec:beaconchain}
We commence the analysis by providing an overview of the size of the network as well as the participation level of the validators in the consensus. 

\subsection{Number of Validators}

\begin{figure}[t]\vspace{-10pt}
\begin{subfigure}[t]{0.43\columnwidth}
    \centering
    \includegraphics[scale=1]{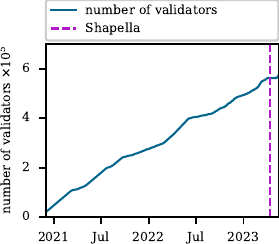}\vspace{-4pt}
        \caption{Number of validators over time. Observe the persistent growth of validators on Ethereum. }
        \label{fig:numbervalidators}\vspace{-2pt}
    \end{subfigure}\hfill
    \begin{subfigure}[t]{0.53\columnwidth}
        \centering
    \includegraphics[scale=1]{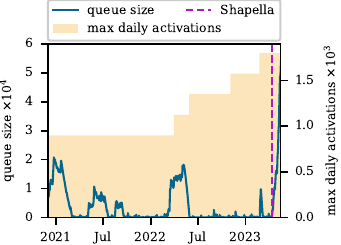}\vspace{-4pt}
    \caption{Number of validators in the queue waiting to join the network along with the maximum number of validators that can be activated each day.}
    \label{fig:queueSize}\vspace{-2pt}
    \end{subfigure}
     \caption{Number of validators (cf. Figure~\ref{fig:numbervalidators}) and number of validators in the queue (cf. Figure~\ref{fig:queueSize}) over time. We mark the Shapella upgrade by the purple dashed line.}\vspace{-2pt}
\end{figure}

Figure~\ref{fig:numbervalidators} displays the total number of validators on the beacon chain. Notice the consistent increase in the number of validators over time. In part, this growth is due to withdrawals only becoming possible after the Shapella upgrade~\cite{HistoryEthereum2023} on 12 April 2023. After the Shapella upgrade, the number of validators remained constant, as both the activation queue (i.e., the queue for validators joining the network) and the exit queue (i.e., the queue for validators exiting the network) were filled and the number of validators that can enter/leave the network is limited. Therefore, the in- and out-flow to the network was constant. Then, almost a month later, on 8 May 2023, the in-flow overtakes the out-flow again, and the number of validators starts to increase again. As of 15 May 2023, there are 572,497 active validators. Each has staked at least 32~ETH on the beacon chain, equivalent to 15.23\% of all Ether in circulation.

We further plot the size of the activation queue in Figure~\ref{fig:queueSize} along with the maximum number of daily activations, which depend on the network size. Post beacon chain launch, an initial rise in the queue size is observable. Subsequent significant increases are noted in mid-2021 and March 2022, which coincide with periods of high Ethereum prices. We observe the most significant and still persistent queue size increase after the Shapella upgrade. As of 15 May 2023, there are 48,903 validators in the queue waiting to join the network, while only 1,800 can join the network each day (yellow area in Figure~\ref{fig:queueSize}). Thus, given these limits imposed on the number of validators joining the network, validators entering a queue of this size will have to wait more than 26 days to become activated. 
\subsection{Proposals}

\begin{figure}[t]
    \centering\vspace{-10pt}
    \includegraphics[scale=1]{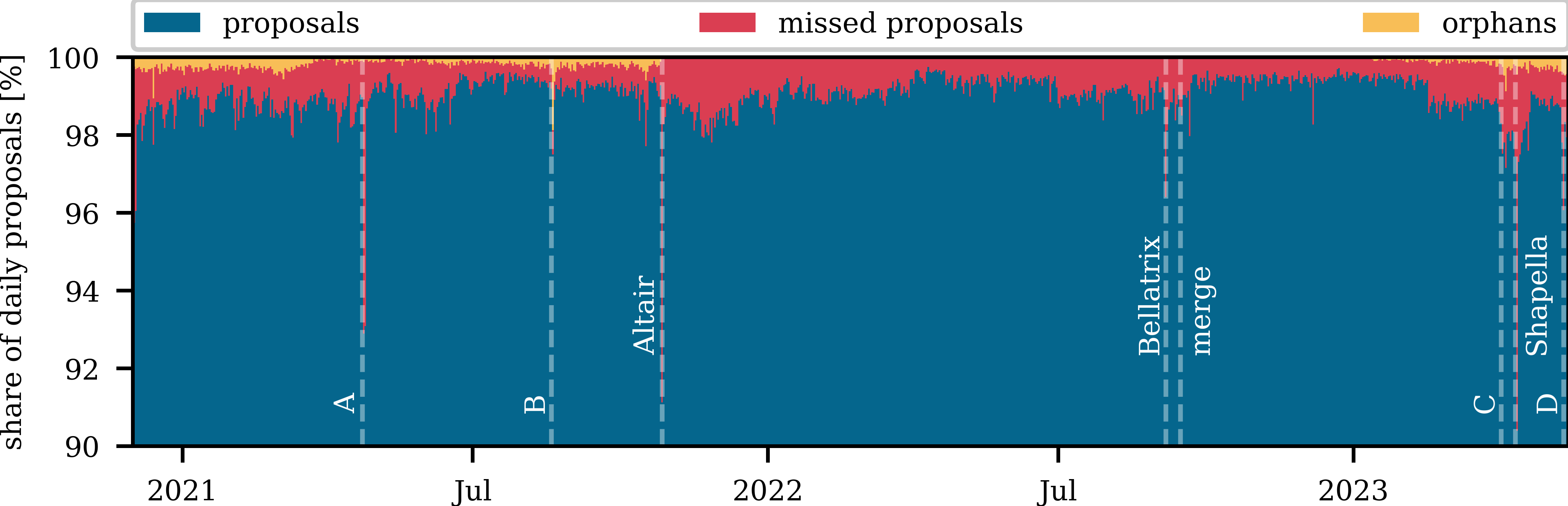}\vspace{-4pt}
    \caption{Daily share of successful (shown in blue), missed (shown in red), and orphaned proposals (shown in yellow). We indicate major upgrades/events that had a noticeable impact on the network participation by white dashed lines. The Altair upgrade heavily reduced the number of orphaned blocks. Also highlighted are incidents A, B, and D. In all three cases bugs in the  Prysm and Teku consensus clients resulted in increased numbers of missed/orphaned proposals. Incident C marks an attack on MEV-Boost.}
    \label{fig:proposals}\vspace{-4pt}
\end{figure}

We commence our analysis of the level of network participation of Ethereum with a longitudinal study of the daily share of successful, missed, and orphaned proposals. In Figure~\ref{fig:proposals}, we visualize the daily share of successful proposals in blue, while we show missed proposals in red and orphaned ones in yellow. The daily number of successful proposals is high throughout the entire history of the beacon chain, with an average success rate of 98.95\%. Some days stand out with significantly lower success rates. Incidents A~\cite{EdgingtonETH2April2021} and B~\cite{EdgingtonETH2August2021} indicate two bugs in the dominant consensus layer (Prysm) that led to an increased missed and orphaned proposals. In particular, incident B was primarily triggered by a service degradation issue with a Lido operator who controlled roughly 2\% of all validators. An existing Prysm bug then exacerbated the situation.

The next sharp increase in the proportion of missed proposals coincides with the Altair upgrade~\cite{HistoryEthereum2023}, which was the first beacon chain upgrade. While the missed proposals reached about 9\% on the day of the upgrade (likely due to operators not updating their clients in time), the proportion of orphaned blocks almost dropped to zero after the update. From then on, the proposal success rate remained relatively stable for almost a year. The Bellatrix upgrade~\cite{HistoryEthereum2023}, which was the second beacon chain upgrade in preparation for the merge, only increased the number of missed proposals on the day of the upgrade. Further, there is no noticeable drop in the proportion of successful proposals during the merge. 

After the merge, we observe a slight increase in the daily share of proposals. Potentially a consequence of the added execution layer rewards received by proposers after the merge --- making block proposals significantly more profitable. However, this trend starts to change in early 2023, with both the share of orphans and missed proposals increasing. In particular, the attack by a validator on MEV-Boost~\cite{MillerPostMortemMEVBoost} (event C) had a lasting impact on the number of missed proposals. After the attack, the timing requirements for validators using MEV-Boost were tightened, which might explain the persistent increase in missed proposals. Shapella~\cite{HistoryEthereum2023}, the third beacon chain upgrade, again increases the number of missed proposals in its immediate aftermath due to validators not updating their clients in time. The final sharp increase in the number of missed proposals during our data collection coincides with the finality issues experienced by Ethereum on 11 and 12 May 2023~\cite{PostMortemFinality2023} as a result of a bug in the Prysm and Teku consensus clients (event D).

We conclude that throughout the entire beacon chain history, proposal participation was very high. Noticeably, the most significant losses in participation, with the exception of upgrades, are at least in part a consequence of bugs in one or more consensus clients. 

\subsection{Attestations}
\begin{figure}[t]\vspace{-10pt}
    \centering
    \includegraphics[scale=1]{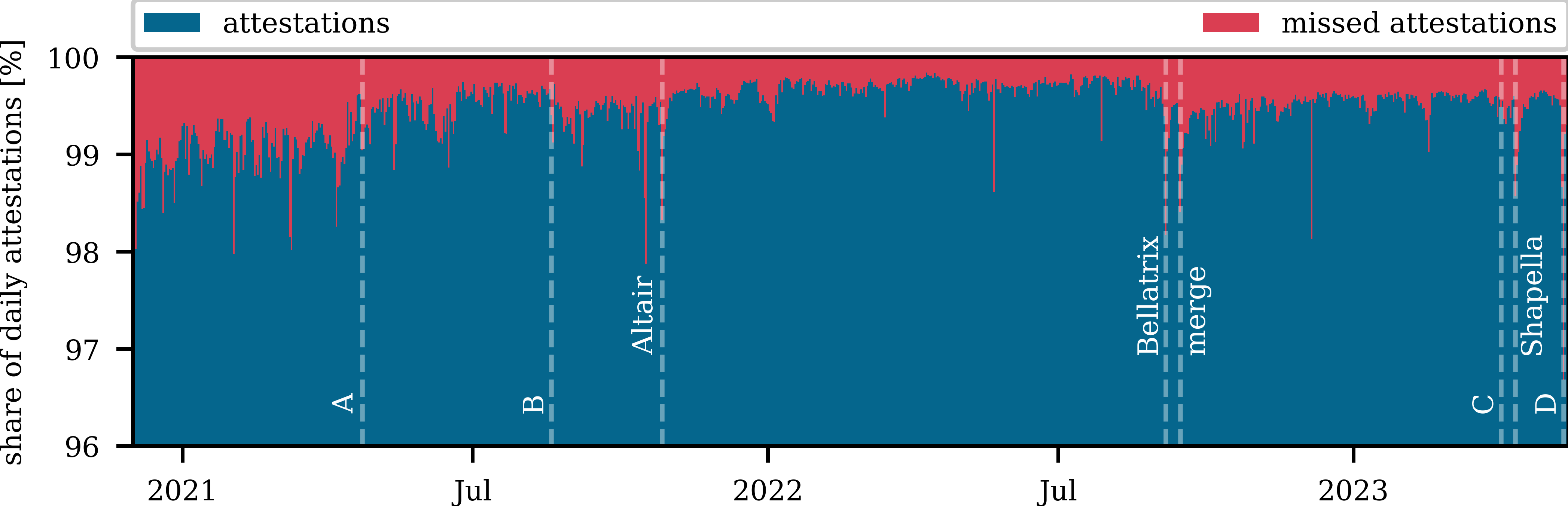}\vspace{-4pt}
    \caption{Daily attestation success rate over time. We indicate major upgrades with a noticeable impact on the network participation by white dashed lines. Incidents A, B, and D, are bugs in the  Prysm and Teku consensus clients that resulted in more missed/orphaned proposals. Further incident C marks an attack on MEV-Boost.}\vspace{-4pt}
    \label{fig:attestations}
\end{figure}

We continue by analyzing the network's participation level for attestations. While lower participation in the block proposals reduces the blockchain's throughput, low participation (i.e., less than two-thirds) in attestation can stall the blockchain's finality. Finality indicates that a block is considered irreversible and permanently added to the blockchain, i.e., it cannot be removed or altered without burning one-third of staked Ether. Thus, high network participation for attestations might even be more crucial than for proposals.

In Figure~\ref{fig:attestations}, we plot the daily share of successful and missed attestations. First, we note that similar to what we previously saw with proposals, the overall participation for attestations is high. On average, 99.46\% attestations are successful each day --- even higher than for proposals. Further, we observe an overall increasing trend in the daily share of successful attestations. By looking at Figure~\ref{fig:attestations} in detail, we notice that incidents A, B, and D, which all mark bugs in one or more consensus clients, are not or less noticeable in the attestation participation levels. All three beacon chain upgrades (i.e., Altair, Bellatrix, and Shapella) and the merge led to short-term increases in the number of missed attestations. In the aftermath of Bellatrix and the merge, the daily share of missed attestations further stays higher than the previous level and only decreases slowly. Importantly, the drop in participation as a result of incident D led to a non-finalizing state. Incident C, the attack of a validator on MEV-Boost, does not appear to have caused a prolonged increase in missed attestations. This is likely because the tightened timing requirements imposed on MEV-Boost validators do not directly affect the attestation procedure.

\subsection{Slashing}

\begin{figure}[t]
\centering\vspace{-10pt}
\begin{minipage}[b]{.48\linewidth}
    \includegraphics[scale=1,right]{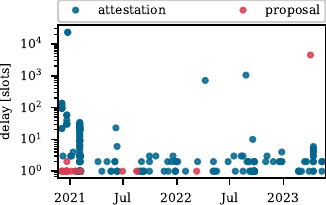}\vspace{7pt}
    \caption{Slashings for attestation violations (shown in blue) and proposal violations (shown in red) over time. We observe a total of 230 attestation violations and a total of 18 proposal violations. The Y axis indicates the delay of the slashing in comparison to the time of the offense.}
    \label{fig:slashing}
\end{minipage}
\hfill
\begin{minipage}[b]{.48\linewidth}
    \includegraphics[scale=1,right]{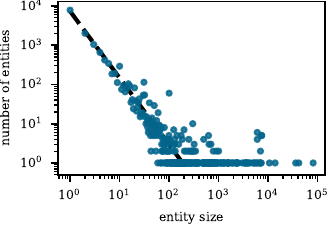}\vspace{-4pt}
    \caption{Distribution of the number of validators in an entity. 7,855 entities are made up of a single validator (top left), while the largest entity consists of 81,165 validators (bottom right). Entity size follows a power-law distribution as demonstrated by the fitted curve. }
    \label{fig:powerlog}
\end{minipage}
\end{figure}

Until now, we focused on network participation. As mentioned previously, validators who do not fulfill their duties will only face minor penalties. To explore the prevalence of serious misbehavior, we examine all slashings --- punishments for serious offenses. Figure~\ref{fig:slashing} plots all slashings for attestation violations (i.e., attesting to a block that ``surrounds'' another or engaging in ``double voting'' by attesting to two candidates for the same block) and for proposal violations (i.e., proposing and signing two different blocks in the same block) from the inception of the beacon chain until 15 May 2023. The Y axis indicates the delay, the number of slots between the slashing, and the offense of the slashing. We find that there are a total of 248 slashings that have taken place: 230 for attestation violations and 18 for proposal violations. Thus, there are only very few (identified) violations in the history of the beacon chain. Additionally, we observe that the vast majority of violations (81.45\%) are identified within ten slots. 

\section{Validator Landscape}\label{sec:validatordecentralization}
Although every validator starts with an equal initial stake of 32~ETH, a single entity can control multiple validators, thereby increasing their \textit{staking power}. In the subsequent analysis, we group validators into entities and examine the level of (de)centralization within the validator landscape by assessing the distribution of staking power amongst entities.  We detail our validator clustering procedure in Appendix~\ref{app:clustering}. Importantly, for liquid staking services where validators are operated by separate entities (i.e., Lido, Stakewise, Swell, and RocketPool), we do not cluster all validators into one entity. Instead, we only cluster validators belonging to staking services into one entity if they are run by the same operator. For example, Lido validators are run by 30 different operators and we thus cluster them by operator in the following. Importantly, it is ambiguous whether the validators associated with a staking service should be classified as a single entity or multiple distinct entities, as we will detail in Section~\ref{sec:pessimistic}. Thus, the following analysis is an \textit{optimistic} analysis of the staking power decentralization, we also present a \textit{pessimistic} analysis in Section~\ref{sec:pessimistic}.

\subsection{Staking Power (De)centralization - Optimistic Vantage Point}

We proceed by analyzing the validator distribution across entities and assess the level of decentralization in staking power over time. To commence, Figure~\ref{fig:powerlog} illustrates the distribution of the entity size --- the number of validators belonging to an entity. We observe that there are many small entities, e.g., nearly 7,855 entities only count a single validator and a few very large entities. The largest entity we identify is Coinbase, with 81,165 validators under their control. There are a couple of entities with the exact same size of around 6,000. These entities are all Lido operators, which are capped in size (cf. Appendix~\ref{app:lidodecen}). In short, we find that the entity size follows a power law distribution as demonstrated by the fitted curve ($\text{number of entities} \propto \text{entity size}^{-\alpha}$). 

\begin{figure}[t!]\vspace{-10pt}
    \centering
    \begin{subfigure}[t]{0.48\columnwidth}
        \includegraphics[scale=1,right]{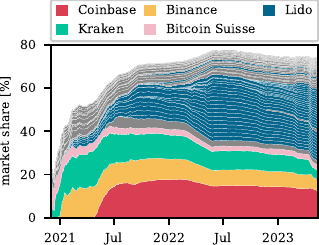}\vspace{-4pt}
    \caption{Staking power distribution of the largest 70 entities as of 15 May 2023 over time.}
    \label{fig:marketshare}
    \end{subfigure}\hfill    
    \begin{subfigure}[t]{0.48\columnwidth}
        \includegraphics[scale=1,right]{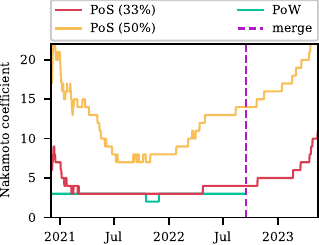}\vspace{-4pt}
    \caption{Comparison between Ethereum PoS (red line) and PoW (green line) Nakamoto coefficient.}
    \label{fig:nakamotovalidators}
    \end{subfigure}\label{fig:decentalizationvalidators}    
    \begin{subfigure}[t]{0.48\columnwidth}
        \includegraphics[scale=1,right]{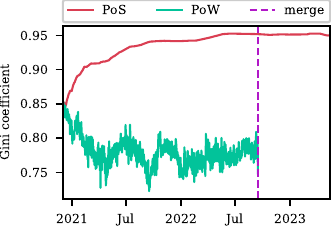}\vspace{-4pt}
    \caption{
    Comparison between  Ethereum PoS (red line) and PoW (green line) Gini coefficient.}
    \label{fig:ginivalidators}
    \end{subfigure}\hfill
    \begin{subfigure}[t]{0.48\columnwidth}
        \includegraphics[scale=1,right]{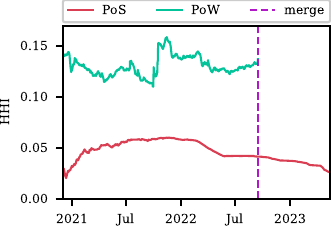}\vspace{-4pt}
    \caption{
    Comparison between  Ethereum PoS (red line) and PoW (green line) HHI.}
    \label{fig:HHIvalidators}
    \end{subfigure}
     \caption{Decentralization analysis for Ethereum PoS in comparison to Ethereum PoW.}\vspace{-3pt}
\end{figure}

We continue by analyzing the (de)centralization of staking power over time. In Figure~\ref{fig:marketshare}, we plot the market share of the biggest 70 entities and highlight the biggest four entities (i.e., Coinbase, Binance, Kraken, and Bitcoin Suisse) as well as the 30 Lido operators. Remarkably, the biggest three entities hold more than 33\% of the staking power from early 2021 until mid-2022. We further note that the combined market share of the 30 Lido operators sits at 32.75\% and all 30 Lido operators are amongst the biggest 70 entities as of 15 May 2023.

To assess the decentralization of the staking power we utilize three measures: \textit{Nakamoto coefficient}, \textit{Gini coefficient}, and \textit{Herfindahl-Hirschman Index (HHI)}, which we provide definitions for in Appendix~\ref{app:decen}.

To start off, the Nakamoto coefficient determines the number of entities that need to be compromised for an adversary to disrupt the blockchain's network~\cite{SrinivasanQuantifying2017}. For Ethereum PoW the adversary requires $>50$\% to disrupt the system, while in Ethereum PoS the requirement is $>33.\bar{3}$\% to stall the system and $>50$\% to break the safety properties~\cite{PoSAttacks2023}). A high Nakamoto coefficient signifies greater decentralization, as it requires a larger number of nodes to be compromised to control the network. Figure~\ref{fig:nakamotovalidators} visualizes the Nakamoto coefficient of Ethereum over time. For Ethereum PoS, we show the Nakamoto coefficient with a $33.\bar{3}$\% threshold and a $50$\%. We further show the Nakamoto coefficient for Ethereum PoW for comparison. Throughout we calculate the decentralization measures for Ethereum PoS based on the staking power and for Ethereum PoW based on a seven-day rolling average of block miners, i.e., mining power.

The Nakamoto coefficient of Ethereum PoW is between two and three from the beginning of the beacon chain to the merge. We notice that for Ethereum PoS, regardless of the threshold, the Nakamoto coefficient is at its highest during the initial phase of the beacon chain and reaches a low point by late 2021. In particular, with the $>33.\bar{3}$\% threshold the Nakamoto coefficient is equal to that of Ethereum PoW for approximately a year.  However, the Ethereum PoS Nakamoto coefficient begins to rise again from 2022 onwards and reaches seven ($>33.\bar{3}$\%) and 20 ($>50$\%) respectively. 

We continue by analyzing the Gini coefficient of Ethereum PoS and comparing it to that of Ethereum PoW (cf. Figure~\ref{fig:ginivalidators}). The Gini coefficient is an inequality measure~\cite{GiniMeasurement1921} whose values range from 0 which indicates perfect equality to 1 which indicates maximal inequality. Interestingly, at the launch of the beacon chain in late 2020, the Gini coefficient of Ethereum PoS and PoW were almost equal at 0.85 --- indicating significant inequality. From then on the Gini coefficients diverge, while that of Ethereum PoW decreases to around 0.77 that of Ethereum PoS increases to 0.95. Thus, the inequality of the Ethereum PoS staking power is significant and even more so exceeds that of Ethereum PoW by a noticeable margin by the time of the merge.

Finally, we calculate the \textit{Herfindahl-Hirschman Index (HHI)}~\cite{RhoadesHerfindahl1993}. The HHI is used for assessing market concentration, i.e., centralization, in economics. Similar to the Gini coefficient, the HHI ranges between 0, indicating a competitive market, and 1, representing a monopolized market with a single dominant firm. Thus, a low HHI value indicates a more decentralized network with numerous independent validators, while a high HHI value points to a more concentrated network, possibly making the system more vulnerable to attacks. Importantly, HHI measures concentration, while the Gini coefficient measures inequality. For example, the Gini coefficient would not distinguish between a single entity with 100\% of the staking power and one with a thousand equal-sized validators. In both cases, the Gini coefficient would be 0 signaling perfect equality. The HHI, on the other hand, is 1 in the first case and 0.001 in the second case.

Figure~\ref{fig:HHIvalidators} plots the HHI of the Ethereum staking power over time and compares it to that of the Ethereum mining power during the PoW era. The green line indicates the HHI of the mining power, which hovers around 0.13 from the beacon chain launch until the merge. For Ethereum PoS, the HHI is significantly lower than that of its PoW counterpart and ranges from 0.02 to 0.06, i.e., Ethereum PoS is less concentrated. The HHI increases initially and then from early 2022 starts to decrease again. The average staking power HHI is  0.046, which from an HHI perspective is equivalent to an industry with 21 equal-sized firms. Thus, while the HHI of 0.046 indicates perfect competition in economics, it is unclear whether such an HHI value is also sufficient to regard the staking power distribution as ``perfectly'' decentralized with 21 equal-sized validators.

We conclude that the decentralization of the staking power of Ethereum is slightly above that of the mining power during PoW times. While Ethereum PoS trails its PoW counterpart in terms of inequality (i.e., Gini coefficient), it fares better in our other decentralization measures (i.e., HHI and Nakamoto coefficient). Recall, that the preceding analysis is an optimistic view as we treat validators associated with the same liquid staking protocol as multiple entities separated by operators. Next, we discuss the validity and impact of this decision.

\subsection{ Staking Power (De)centralization - Pessimistic Vantage Point}\label{sec:pessimistic}

To minimize the risk of operator misbehavior or misconfiguration, multiple liquid staking protocols utilize a permissioned or permissionless set of operators to run the pool's validators instead of putting a single entity in charge of operating all validators. However, while the nodes of liquid staking pools such as Lido are operated by independent operators, there are common incentives and points of failure shared by all validators belonging to the same staking pool. For instance, the smart contracts that operate and govern the liquid staking protocols represent a single point of failure. Flaws in the governance smart contracts could allow an attacker to take over the protocol; similar to the the Tornado Cash hack in May of 2023~\cite{AttackerTornadoCashCrawley2023}. Importantly, this risk is not easily overcome by having users stake across multiple differing implementations of the protocol logic. Not only could these implementations repeat the same mistakes if one does not know where they are in the first place, but the logic itself might also be flawed~\cite{AttackerEulerCrawley2023}. On a different note, governance could be susceptible to additional attacks and the node operators share common incentives. A liquid staking pool that exceeds consensus thresholds can achieve outsized profits in comparison to solo-stakers, for instance through (multi-block) MEV extraction or censorship, and presents an incentive for a liquid staking pool to cartelize the block space~\cite{RiskLSD2022}.

\begin{figure}[t!]\vspace{-10pt}
    \centering
    \begin{subfigure}[b]{0.48\columnwidth}
        \includegraphics[scale=1,right]{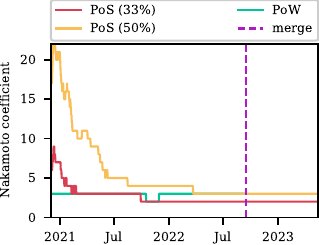}\vspace{-4pt}
    \caption{Comparison between Ethereum PoS (red line) and PoW (green line) Nakamoto coefficient.}
    \label{fig:nakamotovalidatorsw}
    \end{subfigure}\hfill
    \begin{subfigure}[b]{0.48\columnwidth}
        \includegraphics[scale=1,right]{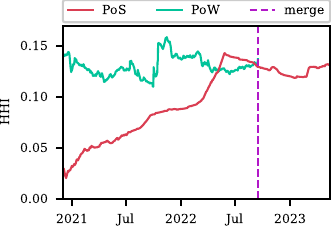}\vspace{-4pt}
    \caption{
    Comparison between  Ethereum PoS (red line) and PoW (green line) HHI.}
    \label{fig:HHIvalidatorsw}
    \end{subfigure}    
    \caption{Decentralization, i.e., Nakamoto coefficient (cf. Figure~\ref{fig:nakamotovalidatorsw}) and HHI (cf. Figure~\ref{fig:HHIvalidatorsw}) of Ethereum PoS and PoW.}\label{fig:decentalizationvalidatorsw}\vspace{-4pt}
\end{figure}

To summarise, there are valid reasons to view the validators associated with a liquid staking protocol as a single entity. Thus, we take this pessimistic view and repeat the staking power decentralization analysis to observe the impact thereof. We compute both the Nakamoto coefficient and the HHI for the Ethereum PoS staking power from this vantage point over time and compare it to that of the mining power distribution of Ethereum PoW (cf. Figure~\ref{fig:decentalizationvalidatorsw}). In both decentralization measures Ethereum PoS initially fares better than its PoW counterpart. However, with time the centralization of the staking power increases and overtakes that of the mining power which stays relatively constant. Interestingly, during the merge, the staking power Nakamoto coefficient with the 50\% threshold equals that of the mining power, i.e., for both mere three entities hold more than 50\%, while the staking power Nakamoto coefficient with the $33.\bar{3}$\% threshold is two. Further, during the execution of the merge, the HHI of staking and mining power were almost equal at around 0.13. From then on, the HHI of the staking power stays relatively stable and sits at 0.13 as of 15 May 2023 ---  equivalent to a landscape with eight equal-sized validators.

Our analysis reveals that the approach to clustering liquid staking pool validators into entities heavily impacts the decentralization of the consensus layer. From the optimistic vantage point, i.e., clustering only validators run by the same operator, the staking power decentralization appears improved in comparison to that of the mining power (cf. Section~\ref{sec:validatordecentralization}). However, from the pessimistic viewpoint, i.e., clustering validators from the same liquid staking pool, the staking power appears as centralized as the mining power used to be. We further point out that even though we take both viewpoints for multiple liquid staking protocols (i.e., Lido, Stakewise, Swell, and RocketPool), Lido is responsible for the vast difference in the staking power decentralization from both vantages. Lido is the largest liquid staking protocol, and its staking power market share has increased throughout the history of the beacon chain to 32.75\% as of 15 May 2023. Lido almost controls a third of all validators: the fraction of validators required to stall Ethereum PoS. Given Lido's significant influence on the network's health, we further provide a detailed discussion of the centralization within Lido in Appendix~\ref{app:lidodecen}. 

\subsection{Impacts of High Centralization}
We continue by discussing the impacts of high centralization. Proposal assignments are known at least one epoch in advance. Thus, entities know ahead of time that they control a continuous sequence of blocks. This peculiarity of Ethereum PoS opens the door to what is known as \textit{multi-block MEV}: value extraction through transaction order manipulation across multiple consecutive blocks. One example of multi-block MEV is oracle manipulation, which becomes cheaper when one is certain to be in control of at least two consecutive blocks~\cite{MackingaTWAP2022}.  

To understand the threat of such attacks under the current staking power distribution, we study the occurrences of uninterrupted block proposal sequences from the optimistic and pessimistic vantage point. An entity only controls a single block in a row in more than 90\% of sequences. Sequences of length two and three both occur multiple times a day --- more frequently from the pessimistic vantage point. Startlingly, regardless of the vantage point, all sequences of length four and longer were controlled by a mere five entities: Kraken, Binance, Bitcoin Suisse, Coinbase, and Lido (solely when considered as one entity). Further, when considered as one entity Lido controls all but one of all sequences of length eight or longer with the longest sequences being of length 13. The certainty for entities to control long sequences opens up additional security concerns for Ethereum PoS that were not present in the same form in Ethereum PoW and exemplify the threats posed by a lack of decentralization in the consensus layer. Further, given the novelty of this attack vector, the possible ramifications are yet to be quantified.
\begin{table}[h]\vspace{-15pt}
    \centering
    \begin{adjustbox}{width=\columnwidth}
    \begin{tabular}{lrrrrrrrrrrrrr}
    \toprule
sequence length & 1 & 2 & 3 & 4 & 5 & 6 & 7 & 8 & 9 & 10 & 11 & 12 & 13 \\
\midrule
occurrences (optimistic) & 5,885,903 & 232,377 & 26,469 & 3,618 & 498 & 79 & 13 & 1 & 0 & 0 & 0 & 0 & 0 \\
occurrences (pessimistic) & 5,345,711 & 388,036 & 73,893 & 17,115 & 4,490 & 1,336 & 382 & 116 & 35 & 18 & 4 & 2 & 1 \\
\bottomrule
\end{tabular}
    \end{adjustbox}
    \caption{Occurrences of continuous proposal sequences by the same entity. }
    \label{tab:sequences}\vspace{-30pt}
\end{table}

\subsection{Performance Advantages of Large Entities}
\begin{figure}[b]\vspace{2pt}
    \centering
    \begin{subfigure}{0.48\columnwidth}
        \includegraphics[scale=1,right]{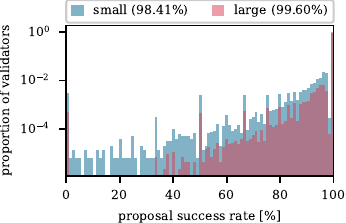}\vspace{-4pt}
    \caption{Successful proposal rate for all validators. On average, the success rate of validators affiliated with large entities is elevated by 1.1\%.}
   
    \label{fig:proposalsuccess}
    \end{subfigure}\hfill
    \begin{subfigure}{0.48\columnwidth}
        \includegraphics[scale=1,right]{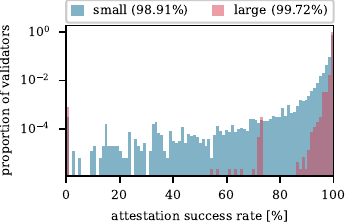}\vspace{-4pt}
     \caption{Attestation success rate across all validators. On average, the success rate of validators affiliated with large entities is elevated by 0.9\%.}
    \label{fig:attestationsuccess}
    \end{subfigure}\vspace{-2pt}
    \caption{Proposal (cf. Figure~\ref{fig:proposalsuccess}) and attestation (cf. Figure~\ref{fig:attestationsuccess}) success rate of all validators split by entity size. We consider all entities with less than 1,000 validators as small entities, and all entities with at least 1,000 validators as large entities.}\label{fig:success}\vspace{-14pt}
\end{figure}

Our preceding analysis demonstrates a dominance of staking services in the validator landscape. Reasons individuals might choose to partake in staking services as opposed to being solo-stakers, could include not only the ease but also the potential of higher returns given the specialized operators. Thus, we investigate whether we observe performance (i.e., the proportion of successful attestations and proposals) differences between small and large entities. In the following, we consider a validator to belong to a small entity if the entity size is smaller than 1,000, else we consider the validator to be part of a large entity. In Figure~\ref{fig:success} we visualize the proposal (cf. Figure~\ref{fig:proposalsuccess}) and attestation (cf. Figure~\ref{fig:attestationsuccess}) success rates of validators by their size.

Notice that overall validators, regardless of their entity size, exhibit very high proposal and attestation success rates with 99.52\% on average for proposals and 99.30\% on average for attestations, i.e., participation is very high regardless of entity size. Still, we observe noticeably higher participation from validators belonging to large entities. This difference in success rates could stem from larger entities being able to afford to use better hardware, having better network connectivity, and having faster emergency response times. We infer that large entities have a small advantage in terms of performance and further note that we provide a more detailed analysis of the performance of large known entities in Appendix~\ref{app:entityperformance}.

\section{Related Work}\label{sec:relatedwork}
\T{Decentralization.} One of the fundamental design principles and objectives of a permissionless blockchain is decentralization. Gencer et al.~\cite{GencerDecentralization2018} were the first to investigate the decentralization of the blockchain consensus layer. Their research revealed that the mining processes of Bitcoin and Ethereum exhibit a significant level of centralization. Subsequent studies by Kiffer et al.~\cite{KifferUnder2021} and Lin et al.~\cite{LinMeasuring2021} for Ethereum PoW also reach the same conclusion. In contrast to these works, we are, to the best of our knowledge, the first to study the decentralization of the staking power in Ethereum PoS. 

\T{Censorship.} In light of the recent imposition of cryptocurrency mixer sanctions by the U.S. Office of Foreign Assets Control, the censorship resilience of Ethereum has come under scrutiny. Wahrstätter et al.~\cite{WahrstatterBlockchain2023} conducted a study on the security impact of blockchain censorship. Their findings indicate an 85\% increase in the inclusion delay of sanctioned transactions and highlight the associated security concerns. Wang et al.~\cite{WangBlockchain2023} specifically focused on the security implications of censoring for validators. While these works focus on the security implications of censorship on the Ethereum blockchain, we measure the consensus layer decentralization of Ethereum. Higher decentralization in the consensus layer is expected to improve censorship resistance.

\T{Maximal extractable value.} MEV is a security concern to the Ethereum consensus~\cite{QinQuantifying2022}. Daian et al.~\cite{DaianFlash2020} and Eskandari et al.~\cite{EskandariSoK2020} provide an early description of MEV. Measurement studies of MEV presented by Torres et al.~\cite{FerreiraFrontrunner2021} and Qin et al.~\cite{QinQuantifying2022} reveal the immense presence and value of MEV. Further, they outline the resulting security risks presented to the consensus layer of a permissionless blockchain. In comparison to these Ethereum PoW studies, we analyze Ethereum PoS and highlight the heightened security risk posed by MEV in PoS. 

\T{Proposer-builder separation.} PBS was designed to decentralize the the consensus layer~\cite{PBSEthereum2023} and its adoption dramatically rose to more than 90\%~\cite{HeimbachEthereum2023,WahrstatterTime2023,YangSoK2022} since the merge. Recently, multiple works have emerged that study the PBS landscape~\cite{Gupta2023Centralizing,HeimbachEthereum2023,SchwarzTime2023,WahrstatterTime2023,YangSoK2022}. These works highlight the increasing trends to centralization of block building under PBS. In this work, we focus on the decentralization of the consensus layer as opposed to block building.

\section{Concluding Discussion}\label{sec:conclusion}

Active participation in and high decentralization of the Ethereum consensus layer are key to the health and security of Ethereum. Thus, understanding the network participation levels in the consensus and the decentralization thereof is essential.

\T{Network participation levels.} Our longitudinal analysis of the consensus participation levels demonstrates the incredibly high participation rates of validators. On average, 99.00\% of blocks are proposed successfully, and 99.53\% of attestations are received. Most dips in participation are during network upgrades or due to problems with consensus clients or large entities. While the vast majority of these temporary drops had no bigger impact on network consensus, the dips in participation level in May 2023, due to bugs in the Prsym and Teku consensus clients, prevented the network from reaching finality for a couple of epochs.

\T{Decentralization.} In our work, we analyze the decentralization of the Ethereum PoS consensus layer. We analyze the decentralization both from an optimistic vantage point, i.e., clustering only validators run by the same operator, and from a pessimistic vantage point, i.e., clustering validators from the same liquid staking pool. Our optimistic analysis demonstrates that the decentralization of the Ethereum staking power exceeds that of the past mining power. However, this does not hold for the pessimistic viewpoint, there the decentralization is approximately equal to that of a landscape with eight equal-sized validators. Lido's growing staking power represents a worry in particular; its staking power is approaching one-third at which point Lido alone could stall the Ethereum network. The Ethereum community is aware of this problem but cannot yet counteract this trend. Even though Ethereum ensures that the hardware requirements for running a validator are low~\cite{PoSvsPoWEthereum2023}, too many people choose the easier route by utilizing staking services instead of running their own validators, i.e., solo staking. As we demonstrate, large entities exhibit better performance compared to small entities, which is likely to cause a disproportionate growth of large entities. Further, staking services utilize the same hardware to run multiple validators and can thus amortize their hardware costs. Finally, the congestion in the activation queue means that it might be too big of a sacrifice for solo-stakers to wait weeks for activation, while the rewards with staking services would be almost immediate. Thus, incentivizing more solo staking is an open problem. 

\T{Security implications.} We further comment on the security implications of a lack of decentralization in the consensus layer --- these go beyond a single party exceeding the consensus threshold in size. MEV is very prevalent on Ethereum and a known risk to the consensus layer, the possibility of multi-block MEV in Ethereum PoS only adds to this. The larger an entity, the higher the chance that it will control multiple consecutive blocks and thereby profit from multi-block MEV. These additional MEV opportunities reserved for larger entities could lead to their market share increasing due to higher expected execution layer rewards.

To conclude, we provide an overview of the Ethereum PoS consensus layer and underwrite the need and desire for increased decentralization --- especially in light of the security concerns posed by heightened consensus layer centralization.

%
%
\bibliographystyle{splncs04}
\bibliography{references}
\newpage
\appendix

\section{Data Collection}\label{app:datacollection}
In the following, we provide a detailed overview of our data collection. We collect beacon chain data (cf. Appendix~\ref{app:beaconchaindata}) by running a Lighthouse node, obtain data to cluster validators from various data sources (cf. Appendix~\ref{app:clusterdata}), and collect Ethereum PoW data (cf. Appendix~\ref{app:datapow}) by running an Erigon node.

\subsection{Beacon Chain Data}\label{app:beaconchaindata}
We collect the majority of our data from the beacon chain with our Lighthouse client and provide an overview of the data collected in the following. 

\T{Validator data.} For each validator that was in the network at some time during our data collection period, we query our Lighthouse client to collect the epoch in which they became eligible for activation (i.e. when they staked 32~ETH in the beacon chain deposit contract), their activation epoch (i.e., when they joined the network), and where applicable the exit epoch (i.e., when they left the network).

\T{Attestations and proposals.} We collect all attestations and proposals assigned to each validator. Then, we record the success status of each. For proposals, we collect data on all validators responsible for a proposal and the success or failure of each block proposal. We further differentiate between missed blocks and orphaned blocks. Orphaned blocks are valid blocks that were proposed but not included in the main chain, as they were outcompeted by another block at the same height. This can occur due to near-simultaneous block proposals, causing temporary chain splits. We use the API provided by \texttt{beaconcha.in} to classify missed proposals~\cite{Beaconchain2023}. Regarding attestations, we gather information on all epoch committees and block attestations. In total, our data set encompasses 62,951,944,703 attestations and  6,447,599 proposals. Out of the 64,385 missed proposals, we identify 6,584 as orphaned.

\T{Slashing.} We analyze all beacon chain blocks to identify any slashings for proposer and attestation violations. For each, we record the slashed validator, the slot in which the slashable offense took place, and the slot in which the proposer was slashed.

\subsection{Validator Clustering Data}\label{app:clusterdata}
As of slot 6,447,598, we have identified 625,193 validators who have participated in at least one attestation duty. For each validator, we collect additional metrics such as labels, deposit addresses, and fee recipient addresses. This information aids us in clustering these validators into distinct entities in subsequent analysis steps.
 
\T{Deposit addresses.} Deposit addresses are those that provide the 32~ETH required to fund a validator. Note that a validator can be associated with multiple deposit addresses. We collect all such addresses by monitoring the logs from the beacon chain deposit contract. In total, we identified 99,711 distinct deposit addresses in our dataset. 

\T{Fee recipient addresses.}
A fee recipient address is an Ethereum address specified by the validator to collect the execution layer rewards for their block proposals. Validators sharing the same fee recipient address are likely to represent the same entity. To identify these addresses, we extract the fee recipient field for each block from our Erigon client. If a block is not created through \textit{proposer-builder separation (PBS)}~\cite{PBSEthereum2023}, this address corresponds to the validator's fee recipient address. Otherwise, it belongs to the block builder. Since a significant number of blocks are constructed via PBS, we also inspect the last transaction in each block. If the block was built through PBS, the builder transfers the fees to the validator in this transaction~\cite{MEVBoostGit2023}. Hence, if no such transaction exists, we record the block's fee recipient address as the validator's fee recipient address. If such a transaction is present, we record the receiver of the ETH in the last transaction as the validator's fee recipient address. Our dataset identifies 11,515 unique fee recipient addresses.

\T{Validator labels.} A validator label is a name that associates a validator with an entity, e.g., staking pools, staking-as-a-service providers, centralized exchanges, or institutional validators. Not all validators have an associated label, as some operators may choose not to disclose their identity or affiliation. We obtain and combine validator labels from the Rated Network API~\cite{Rated2023}, through scraping data from \texttt{beaconcha.in}~\cite{Beaconchain2023} and through manual identification of deposit patterns for Coinbase (cf. Appendix~\ref{app:coinbase}). 

\T{Address labels.}
We further obtain address labels for deposit addresses through data scraping from Etherscan~\cite{Etherscan2023}. Specifically, we collect \textit{Ethereum Name Service (ENS)} domain names, i.e., web3 usernames~\cite{ENS2023}. We will utilize address labels in addition to validator labels to identify and label entities.

\T{Etherclust data set.}
We cluster Ethereum addresses through the reused centralized exchange deposit addresses method introduced by Victor~\cite{VictorAddress2020}. To be precise, we cluster together addresses that utilize the same exchange deposit address. The clustering compresses  6,788,215 addresses into 1,410,523 entities with more than one address -- representing approximately 2.93\% of the 231,625,425 unique Ethereum addresses as of 15 Mai 2023~\cite{EtherscanUniqueAdresses2023}.

\subsection{Ethereum Proof-of-Work Miners}\label{app:datapow}
We amend our analysis of the Ethereum PoS consensus (de)centralization by comparing it to the former PoW consensus. Thus, we collect the miners for each block from the first block of the Ethereum blockchain on 30 July 2015 to block 15,537,392 (i.e., the last block before the merge on 15 September 2022) from our Erigon client. Additionally, we obtain labels for the miner addresses from Etherscan~\cite{EtherscanMiningAccounts2023}.

\section{Entity Clustering}\label{app:clustering}
We provide a detailed overview of our entity clustering procedure in the following. Our clustering relies on the data set described in Section~\ref{app:clusterdata}. Barring exceptions, which we detail in Appendix~\ref{app:entitycluster}, the clustering process is executed in four steps:
\begin{enumerate}[topsep=0pt,itemsep=0pt]
\item Validators sharing the same label (i.e., validator labels obtained from Rated Network API and \texttt{beaconcha.in}) are grouped into a single entity.
\item Entities or individual validators with at least one common \textit{deposit address} (i.e., the address(es) used to stake the 32~ETH) are merged into the same entity.
\item Entities or individual validators sharing at least one \textit{fee recipient address} are consolidated into the same entity if at least one of the entities does not have a label yet and has consistently used the same \textit{fee recipient address}. This is a conservative approach that reduces the risk of incorrect clustering.
\item Entities or individual validators whose \textit{deposit address(es)} belong to the same entity, as per the Etherclust data set, are combined.
\end{enumerate}

We note that the inherent complexities and nuances to consider during entity clustering, especially given the variety of strategies and structures present in the Ethereum staking landscape, necessitate some manual adjustments and informed exceptions to our clustering process. Throughout, we always opt for the conservative route to maintain the integrity of our analysis. We detail our Coinbase validator identification process (cf. Section~\ref{app:coinbase}) and all clustering exceptions (cf. Section~\ref{app:entitycluster}) next.

\subsection{Coinbase Validator Labels}\label{app:coinbase}
We undertake an additional identification process for Coinbase validators, as especially recent ones have not been labeled by our data sources, i.e., Rated Network API~\cite{Rated2023} and \texttt{beaconcha.in}~\cite{Beaconchain2023}. Coinbase adheres to a unique and easily identifiable pattern when rolling out new validators, allowing a straightforward detection of their validators. In particular, the following properties hold for any Coinbase validator: 
\begin{enumerate}[topsep=0pt,itemsep=0pt]
    \item Each validator employs a unique deposit address.
    \item The Ether sent to the deposit address originates from a Coinbase address.
    \item Post-deployment, the surplus Ether is redirected back to a Coinbase address.
    \item A small nominal amount of Ether, typically around 0.0006~ETH, is left in the deposit address. 
\end{enumerate}

Thus, we run through all unlabeled validators and label those for which all of the above properties hold as Coinbase. In doing so, we label an additional 3,222 validators as Coinbase. Note that we obtain a list of all Coinbase addresses from Etherscan~\cite{EtherscanCoinbaseAccounts2023}.

\subsection{Entity Clustering Exceptions}\label{app:entitycluster}
In the following, we detail the primary challenges and the corresponding adjustments we make during entity clustering. These exceptions allow us to mitigate any unnecessary over-clustering.

\T{Validator label exclusion.} 
Our data set includes labels for companies providing only the hardware and software for staking, with the staker controlling the deposit and fee recipient addresses --- \textit{non-custodial staking}. In such cases, we aim to sidestep clustering a validator with the company providing the underlying resources. 

Instead, we strive to cluster it with other validators controlled by the same staker. The distinction between custodial and non-custodial staking services is not always apparent and some providers even changed over time. We do not cluster an entity by its validator label, if all three of the following criteria are met: (1) the service permits non-custodial staking, (2) the staker knows their exact validator ID, and (3) upon clustering all labels of this entity together, the entity exhibits numerous used deposit addresses and at least one of these deposit addresses is associated with an unrelated address label (ENS name), suggesting user control over this address.

We remove these validator labels for non-custodial staking services that meet all our previously outlined criteria. Namely, these non-custodial staking services are \textit{Staked.us}, \textit{Stakefish}, and \textit{Bloxstaking}.

\T{Validator label-only clustering.}
Liquid staking protocols such as \textit{Lido} and \textit{StakeWise}, multiple different operators run the protocol's validators. The validator labels we obtain from the Rated Network API allow us to identify which validators are run by which operator. Given all validators belonging to these protocols, regardless of the operator running them, they typically share the same deposit addresses. Our standard clustering method would cluster all validators belonging to such a liquid staking protocol as one entity. 

To retain the higher resolution provided by the validator labels obtained from the Rated Network API, we modify our approach for validators from \textit{Lido} and \textit{StakeWise} and only focus on their validator labels during the clustering process. Throughout our analysis we will, at times, analyze the impact of viewing all validators belonging to a liquid staking service as one entity as opposed to multiple entities depending on the operators in control of the validators.

\T{Fee recipient address exclusion.}
Node operators participating in non-custodial staking services (e.g., Rocket Pool), have the option to designate a communal smoothing pool address as their fee recipient address. This shared pool is designed to even out the potentially fluctuating MEV rewards, which can be particularly beneficial for smaller node operators. However, this setup causes all Rocket Pool nodes that join the smoothing pool to be clustered as a single entity, even though they are controlled by multiple different operators.

To counteract this, we exclude certain fee recipient addresses from our clustering process. These include fee recipient addresses associated with \textit{Stakefish}, \textit{Staked.us}, \textit{Ethpool Staking}, \textit{Ankr}, and the \textit{Rocket Pool smoothing pool}. A complete list of these excluded addresses can be found in Table~\ref{tab:excludedFRA}.
\newcommand{\specialcell}[2][c]{%
  \begin{tabular}[#1]{@{}c@{}}#2\end{tabular}}
  
\begin{table*}[ht]
\centering
    \begin{adjustbox}{width=0.8\linewidth}
        \begin{tabular}{@{}lr@{}} 
            \toprule
            & fee recipient address \\
            \midrule
            Stakefish &\specialcell{\texttt{0xffee087852cb4898e6c3532e776e68bc68b1143b}  \\
\texttt{0x54cd0e6771b6487c721ec620c4de1240d3b07696} \\
\texttt{0x5caf7c1b096cf684b09ece3d3a142db0d46fc58e} \\
\texttt{0xe94f1fa4f27d9d288ffea234bb62e1fbc086ca0c} } \\ \midrule
            Rocket Pool Smoothing Pool & \texttt{0xd4e96ef8eee8678dbff4d535e033ed1a4f7605b7} \\ \midrule
            Ethpool & \texttt{0xb364e75b1189dcbbf7f0c856456c1ba8e4d6481b} \\ \midrule
            MEV Builder & \texttt{0xac7ea48093b61f2e217b9d077d69d9d55ca1b106} \\ \midrule
            Ankr & \specialcell{\texttt{0x3bef77233e52d23969958587127d99ec2367c2bd}  \\
\texttt{0x90b0c836a19a74195d45fad2d2d3895a7a3eab08} \\
\texttt{0x6a0db4cef1ce2a5f81c8e6322862439f71aca29d}  }  \\ 
            \bottomrule
        \end{tabular}
    \end{adjustbox}
    \caption{Fee recipient addresses which we exclude during clustering and which entity they are associated with. These addresses would lead to unwanted linking of entities.}
    \label{tab:excludedFRA}\vspace{-8pt}
\end{table*}

\T{Etherclust exceptions.} In step four of our clustering procedure (cf. Appendix~\ref{app:clustering}), we further manually ensure that entities unlikely to belong to the same entity are not further merged. To be precise, we refrain from clustering well-known entities with any other entity. We make this exception on five separate occasions: (1)~multiple LIDO operators, (2)~multiple Rocket Pool operators, (2)~Bitfinex, Binance, and Whale 0xEAB8, (3)~Coinbase and zachrellim.eth, and (4)~Stakely.io and StaFi.

\section{Decentralization Measures}\label{app:decen}
In the following, we provide definitions of the three decentralization measures used in our analysis.
\T{Nakamoto Coefficient}
The Nakamoto coefficient is a measure used to assess the decentralization of blockchain. In more detail, the Nakamoto coefficient represents the number of independent entities needed to disrupt the blockchain. For PoW blockchain, generally >50\% of the mining power is required to disrupt the network. Thus, the Nakamoto coefficient is the minimum number of independent entities that hold >50\% of the mining power together. For Ethereum PoS $>33.\bar{3}$\% of the staking power is required to stall the network. While this causes problems for the network, it can recover from this by slashing inactive validators. With more than $>50$\%, on the other hand, the attacker can dominate the fork choice algorithm and honest validators would eventually follow suit~\cite{PoSAttacks2023}.

\T{Gini Coefficient}
The \textit{Gini coefficient} is an inequality measure. Mathematically, the Gini coefficient is calculated as follows
\begin{equation*}
G= \frac{\sum _{i=1}^n \sum _{j=1}^n |x_i-x_j|}{2n^2 \bar{x}}
\end{equation*}
where $n$ is the number of entities, and $x_i$ denotes the wealth of person $i$. A value of 0 indicates perfect equality, whereas a value of 1 indicates complete inequality. Note that in the context of Ethereum, we take a validator's market share as their wealth. A low Gini coefficient then indicates that the wealth is equally distributed amongst validators.

\T{Herfindahl-Hirschman Index}
The \textit{Herfindahl-Hirschman Index (HHI)} is used for assessing market concentration and competition in economics. Mathematically, it is expressed as
\begin{equation*}
HHI = \sum_{i=1}^{n} s_i^2,
\end{equation*}
where $n$ is the number of entities in the market, and $s_i$ denotes the market share of entity $i$ as a fraction. Thus, the HHI ranges between 0, indicating low concentration, and 1, representing a concentrated market. In the context of Ethereum and staking, we apply the HHI to analyze validator concentration and potential centralization risks. For us, $n$ is the number of entities, and $s_i$ is the proportion of staking power controlled by entity $i$. A low HHI value indicates a more decentralized network with numerous independent validators, while a high HHI value points to a more concentrated network, possibly making the system more vulnerable to attacks.

\section{Entity Performance}\label{app:entityperformance}

In the following, we take an in-depth look at the participation rate of the five biggest  entities (Lido, Coinbase, Kraken, Binance, and Bitcoin Suisse) in Table~\ref{tab:bigentities}. Note that, here, we consider Lido as one entity but provide an overview of the performance of the 30 individual operators in Appendix~\ref{app:lidoperator}. From Table~\ref{tab:bigentities}, we can see that all big operators have very high and similar participation rates for attestations (higher than 99.80\% for all) and proposals (higher than 99.44\% for all). Further, it is not clear which of the five biggest entities has the highest participation rate. While Coinbase has the highest attestation success rate, Lido has the highest proposal success rate. 

\begin{table*}[h]\vspace{-10pt}
\centering
    \begin{tabular}{@{}lrr@{}}
\toprule
 &  attestation success rate [\%]  &  \hspace{0.5em}proposal success rate [\%] \\
\midrule
Lido & 99.858 & 99.755 \\
Coinbase & 99.891 & 99.485 \\
Binance & 99.877 & 99.448 \\
Kraken & 99.821 & 99.587 \\
Bitcoin Suisse & 99.810 & 99.529 \\
\bottomrule
\end{tabular}\vspace{4pt}
    \caption{Attestation and proposal success rate for the five biggest entities.}
    \label{tab:bigentities}\vspace{-30pt}
\end{table*}   

\subsection{Lido Operator Performance Comparison}\label{app:lidoperator}
We further take an in-depth look at the participation metrics across all 30 Lido staking operators in Table~\ref{tab:lidooperators}. A pattern of consistently high performance emerges throughout their operational history. However, RockLogic GmbH and Chorus One stand out with a comparatively lower performance.
The dip in performance for RockLogic GmbH might be attributed to a slashing incident that occurred on 13 April 2023. The incident was triggered by the inadvertent duplication of validator keys across two active clusters, which resulted in a double vote and consequent slashing of eleven validators~\cite{PostMortemLidoRockLogic2023}.
As for Chorus One, an extended downtime event in October 2021 might be the origin of this reduced performance. The downtime was accidental and was the result of complications during node migrations~\cite{PostMortemLidoChorus2023}.

\begin{table*}[t]
\centering
    \begin{tabular}{@{}lrr@{}}
        \toprule
         & attestation success rate [\%] & \hspace{0.5em}proposal success rate [\%] \\
        \midrule
        Allnodes Lido & 99.981 & 99.932 \\
        Kukis Global Lido & 99.973 & 99.971 \\
        Sigma Prime Lido & 99.967 & 99.911 \\
        Attestant Lido & 99.964 & 99.865 \\
        Everstake Lido & 99.955 & 99.829 \\
        Blockdaemon Lido & 99.952 & 99.806 \\
        ChainLayer Lido & 99.952 & 99.915 \\
        RockX Lido & 99.951 & 99.822 \\
        P2P.ORG - P2P Validator Lido & 99.945 & 99.754 \\
        HashQuark Lido & 99.943 & 99.856 \\
        Stakely Lido & 99.943 & 99.892 \\
        InfStones Lido & 99.942 & 99.660 \\
        CryptoManufaktur Lido & 99.932 & 99.942 \\
        Staking Facilities Lido & 99.925 & 99.860 \\
        Prysmatic Labs Lido & 99.912 & 99.768 \\
        Blockscape Lido & 99.905 & 99.882 \\
        Anyblock Analytics Lido & 99.897 & 99.824 \\
        Nethermind Lido & 99.894 & 99.718 \\
        Kiln Lido & 99.886 & 99.705 \\
        Simply Staking Lido & 99.852 & 99.728 \\
        Figment Lido & 99.835 & 99.842 \\
        ConsenSys Codefi Lido & 99.822 & 99.791 \\
        Stakefish Lido & 99.803 & 99.833 \\
        DSRV Lido & 99.784 & 99.914 \\
        Stakin Lido & 99.776 & 99.813 \\
        Certus One Lido & 99.765 & 99.430 \\
        ChainSafe Lido & 99.750 & 99.768 \\
        BridgeTower Lido & 99.697 & 99.788 \\
        Chorus One Lido & 99.489 & 99.050 \\
        RockLogic GmbH Lido & 99.396 & 98.820 \\
        \bottomrule
        \end{tabular}\vspace{4pt}
    \caption{The 30 Lido staking operators exhibit consistently high performances in both attestation and proposal success rates.}
    \label{tab:lidooperators}\vspace{-10pt}
\end{table*}

\section{Lido (De)centralization}\label{app:lidodecen}
\begin{figure}[t]\vspace{-10pt}
    \centering
    \begin{subfigure}[t]{0.48\columnwidth}
        \includegraphics[scale=1,right]{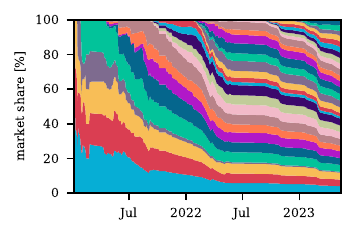}
    \caption{Lido staking power distribution across Lido operators over time.}
    \label{fig:lidomarketshare}
    \end{subfigure}\hfill
    \begin{subfigure}[t]{0.48\columnwidth}
        \includegraphics[scale=1,right]{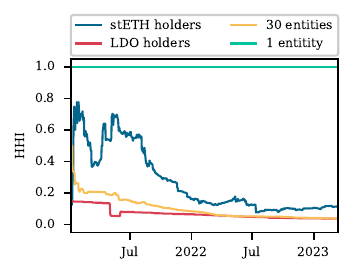}
    \caption{Concentration (HHI) of Lido's stETH tokens, LDO tokens, and the 31 Lido staking operator entities included in our data set.}
    \label{fig:lido}
    \end{subfigure}
     \caption{ Decentralization analysis (i.e., staking power distribution (cf. Figure~\ref{fig:lidomarketshare}) and HHI of staking power distribution (cf. Figure~\ref{fig:lido})) for Lido over time.}\label{fig:todo}
\end{figure}

As the dominant liquid staking protocol with a significant influence on the network's health, Lido merits a closer look. In the following, we provide an in-depth analysis of the distribution of power within Lido. As of the time of writing, Lido operates through 30 permissioned node operator companies tasked with performing the staking. Figure~\ref{fig:lidomarketshare} visualizes the Lido staking power distribution across these operators. We notice that the number of operators is increasing with time and that at the end of our data collection window, the 30 operators are almost all of the same size with around 6,000 validators each.

Importantly, those operators alone do not have the power to arbitrarily change the protocol. Currently, Lido's operations on Ethereum are governed by LDO token holders through an Aragon DAO. This governance encompasses a wide range of aspects, including the Lido treasury, staking withdrawal keys, the registry of node and oracle operators, DAO Access Control List permissions, and the execution of EVM scripts. In effect, LDO holders have root access to the Lido protocol. Proposals are being considered to circumscribe the DAO's authority by enabling stETH holders having the power to veto certain decisions. However, a wide and diverse distribution of the LDO and stETH tokens is indispensable for a healthy protocol.

Figure~\ref{fig:lido} visually represents the concentration trends among LDO and stETH holders over time, as measured by the HHI. It is encouraging to observe a progressive decentralization, with stETH holders' HHI approximating 0.1 and LDO holders’ HHI around 0.034 towards the end of the observed period. Similarly, Lido’s staking operators exhibit an ongoing dispersion trend with a current HHI of 0.036. Overall, the concentration has been on a decline for the past two years. Nonetheless, it is imperative to recognize and address risks the current Lido dominance poses to the Ethereum network, as for instance, the Lido smart contracts could be a single point of failure.

\section{Ethereum Proof-of-Work (De)centralization}\label{app:powDecen}
\begin{figure}[ht]\vspace{-6pt}
    \centering
    \begin{subfigure}[t]{0.48\columnwidth}
        \includegraphics[scale=1,right]{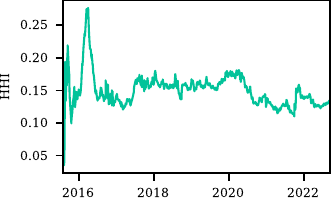}
    \caption{HHI, i.e., concentration, of Ethereum PoW consensus from Ethereum genesis until the merge.}
    \label{fig:HHIvalidatorspow}
    \end{subfigure}\hfill
    \begin{subfigure}[t]{0.48\columnwidth}
        \includegraphics[scale=1,right]{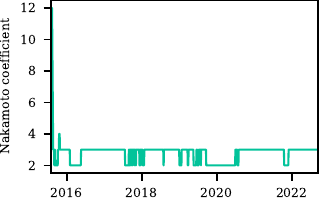}
    \caption{Nakamoto coefficient of Ethereum PoW consensus from Ethereum genesis until the merge.}
    \label{fig:nakamotovalidatorspow}
    \end{subfigure}\label{fig:decentalizationvalidatorspow}
     \caption{Decentralization, i.e., HHI (cf. Figure~\ref{fig:HHIvalidatorspow}) and Nakamoto coefficient (cf. Figure~\ref{fig:nakamotovalidatorspow}) of Ethereum PoW.}
\end{figure}

In the following, we provide some additional insight regarding the network (de)centralization fluctuations during Ethereum's PoW era. Figures~\ref{fig:HHIvalidatorspow} and~\ref{fig:nakamotovalidatorspow} illustrate an early surge in network decentralization following Ethereum's launch. However, this level of decentralization quickly diminished and largely maintained steady levels until the end of PoW.

\end{document}